\documentclass[aps,prb,groupedaddress,amsmath,twocolumn]{revtex4}
    \usepackage{bm}
    \usepackage{graphicx}
    \usepackage{multirow}

    \newcommand\beq{\begin{equation}}
\newcommand\eeq{\end{equation}}
\newcommand\beqa{\begin{eqnarray}}
\newcommand\eeqa{\end{eqnarray}}

\begin{document}

\title{A branch-point approximant for the equation of state of hard
spheres}
\author{Andr\'es Santos}
\email{andres@unex.es}
\homepage{http://www.unex.es/eweb/fisteor/andres/}
\author{Mariano L\'opez de Haro}
\email{malopez@servidor.unam.mx}
\homepage{http://xml.cie.unam.mx/xml/tc/ft/mlh/}
\thanks{on sabbatical leave from Centro de Investigaci\'on en
Energ\'{\i}a, Universidad Nacional Aut\'onoma de M\'exico
(U.N.A.M.), Temixco, Morelos 62580, M{e}xico}
\affiliation{Departamento de F\'{\i}sica, Universidad de
Extremadura, E-06071 Badajoz, Spain}
\date{\today}

\narrowtext
\begin{abstract}
Using the first seven known virial coefficients and forcing it to
possess two branch-point singularities, a new equation of state for
the hard-sphere fluid is proposed. This equation of state predicts
accurate values of the higher virial coefficients, a radius of
convergence smaller than the close-packing value, and it is as
accurate as the rescaled virial expansion and better than the Pad\'e [3/3]
equations of state. Consequences regarding the convergence
properties of the virial series and the use of similar equations of
state for hard-core fluids in $d$ dimensions are also pointed out.
\end{abstract}

\maketitle

\section{Introduction}
The virial expansion of the equation of state (EOS) is an expansion
in powers of (usually) the number density $\rho$ that was originally
introduced phenomenologically by Kammerlingh Onnes\cite{KO909} in
1909 in order to provide a mathematical representation of
experimental data. Later, in what may be considered as one of the
great achievements in statistical physics in the twentieth century,
Mayer\cite{MM40} was able to derive such an expansion for the
pressure $p$ of a classical fluid in terms of its density. The
corresponding virial coefficients (usually denoted by $B_j$) turn
out to be related to integrals over the interaction among groups of
fluid particles and are in general functions of the absolute
temperature $T$. In the case of hard-sphere (HS) fluids, which are
the subject of this paper, the virial coefficients are however
independent of $T$. In particular, the value of the second virial
coefficient for HSs in $d$ dimensions is $B_2=2^{d-1}v_d
\sigma^d$, where $\sigma$ is the diameter of the spheres and
$v_d=(\pi/4)^{d/2}/\Gamma(1+d/2)$ is the volume of a $d$-dimensional
sphere of unit diameter, a result first derived for
three-dimensional HSs ($d=3$) by van der
Waals.\cite{vdW899} Analytical expressions for $B_3$ and $B_4$ are
also available in the
literature\cite{J896,B896,vL899,B899,T36,R64,H64,LB82,BC87,CM04a,L05}
but higher virial coefficients must be computed numerically and,
since this represents a non trivial task, up to now only values up
to the tenth virial coefficient have been
reported.\cite{MRRT53,RR54,RH64a,RH64b,R65,RH67,KH68,K76,K77,K82a,K82b,vRT92,vR93,VYM02,CM04b,LKM05,CM05,KR06,CM06,BCW08}

The virial expansion for $d$-dimensional HS systems is often
expressed in terms of the packing fraction $\eta$ defined as
$\eta=v_d \,\rho \,\sigma^d$. Hence, for these systems the
compressibility factor $Z\equiv p/\rho k_B T$ (with $k_B$ the
Boltzmann constant) is given by

\beq \label{virial2} Z(\eta)=1+\sum_{j=2}^\infty b_j \eta^{j-1},
\eeq where the (reduced) virial coefficients $b_j\equiv B_j/(v_d
\sigma^d)^{j-1}$ are now pure numbers.

The availability of only a few virial coefficients represents a
restriction on the usefulness of the virial expansion and many
issues about it are still unresolved. For instance, its radius of
convergence is not known eventhough lower bounds are
available.\cite{LP64,FPS07} Secondly, although all the available
virial coefficients in  $d=2$, $d=3$, and $d=4$ are positive,
even the character of the series (either alternating or not) is
still unknown. In fact results from higher dimensions suggest that
the positive character might not be true for the higher virial
coefficients of hard disks and spheres.\cite{CM06,RMHS08,AKV08}
Finally, people have usually recurred to  approximate EOSs obtained
through the knowledge of the limited number of virial coefficients
via various series acceleration methods such as Pad\'e or Levin
approximants. However, the expectation that these EOSs would
ultimately lead to the complete phase behavior of the system has not
been fulfilled. Hence, the question of whether the virial series
contains relevant information related to the phase behavior of the
HS system also remains as an open one.

Recently it has been clearly established that the EOSs for hard
hyperspheres ($d\geq 4$) predicted by the Percus--Yevick (PY)
integral equation possess a branch-point singularity on the negative
real axis that is responsible for the radius of convergence and the
alternating character of the virial series.\cite{RMHS08,AKV08} It is
very likely that these features are not artifacts of the PY
approximation but would be shared by the exact EOSs. However, in the
case of hard spheres ($d=3$), the radius of convergence of the PY
EOS is artificially $\eta=1$ and, as stated above, there is no
definite indication about the nature of the singularity responsible
for the true radius of convergence or its value.\cite{CM06}

The main aim of this paper is to shed some more light on the
character of the virial series of the three-dimensional HS fluid.
The idea is to propose a new (heuristic) EOS for HS systems in $d$
dimensions that, for reasons that will become clear later, we will
refer to as a `branch-point approximant.' Such a proposal is not
geared specifically towards obtaining an accurate EOS but rather
relies on the notion that the radius of convergence of the virial
series might be dictated by a branch-point singularity. In any case,
the plausibility of this notion will be assessed by comparing the
predictions of high virial coefficients coming out of the proposal
both with the exact values of these coefficients for each $d$ and
with the performance of other proposals for the EOS (rescaled virial
expansions and Pad\'e approximants).

The paper is organized as follows. In the next section we introduce
the new EOS including a branch-point singularity and examine the
case of three-dimensional HSs. Section \ref{sec2} refers to
the use of the same type of EOS for different dimensionalities. We
close the paper in Sect.\ \ref{sec3} with some discussion and
concluding remarks.

\section{The case of hard spheres ($d=3$)}
\label{sec1}
We  begin by proposing a `branch-point approximant' for the EOS of
$d$-dimensional HS systems, namely
\beq
Z(\eta)=1+\frac{
1+c_1\eta+c_2\eta^2+c_3\eta^3-\left(1+2a_1\eta+a_2\eta^2\right)^{3/2}}{A(1-\eta)^k},
\label{1}
\eeq
where $A$, $a_1$, $a_2$, $c_1$, $c_2$,
 $c_3$, and $k$  are parameters to be determined. This functional form (with $k=3$) is
inspired by the EOS for hard hyperspheres in $d=5$ predicted by the
PY theory through the virial route.\cite{FI81,L84,GGS90,BMC99} As stated above, we will assume
the approximant form given in Eq.\ \eqref{1} for three-dimensional
HSs as a toy model to highlight the possibility that the radius of
convergence of the virial series in this system might be dictated by
a branch-point singularity. According to the philosophy of an
approximant,  the six coefficients $A$, $a_1$, $a_2$, $c_1$, $c_2$,
and
 $c_3$ are obtained from the knowledge of the  virial
 coefficients $b_2$--$b_7$. The resulting expressions are given in
Table \ref{tab1}, where we have called
\beq
S_{k,n}\equiv \sum_{j=2}^n\binom{k}{n-j}(-1)^j b_j.
\label{8}
\eeq
\begin{table}
\caption{\label{tab1} Expressions (for general $k$) and numerical
values (for $k=d=3$) of  $a_1$, $a_2$, $A$, and $c_1$--$c_3$.}
\begin{ruledtabular}
\begin{tabular}{ccc}
Coefficient & Expression&  Value \\
\hline
$a_1$&${S_{k,6}}/{S_{k,5}}$&$0.271232$\\
$a_2$&$7a_1^2-6{S_{k,7}}/{S_{k,5}}$&$1.94804$\\
$A$&$\frac{3}{8}{( a_2-a_1^2)^2}/{S_{k,5}}$&$1.51486$\\
$c_1$&$ 3 a_1+b_2 A$&$6.87314$\\
$c_2$&$\frac{3}{2} ( a_2+a_1^2)-S_{k,3}A$&$0.00268343$\\
$c_3$&$\frac{1}{2}a_1 ( 3 a_2- a_1^2)+S_{k,4} A$&$1.33515$\\
\end{tabular}
\end{ruledtabular}
\end{table}

Although the choice for $k$ is in principle arbitrary, a natural
one seems to take $k=d$. Hence, in this Section we assume $k=3$. A
special situation takes place if $a_2=a_1^2$. In that case, the
denominator ($S_{3,5}$) in the expression for $A$ must vanish in
order to have a finite value, i.e., $b_5=b_2 - 3 b_3 + 3 b_4$. Since
this denominator also appears in the expressions for $a_1$ and
$a_2$, the respective numerators ($S_{3,6}$ and $S_{3,7}$) must also
vanish, i.e., one must have $b_6= 3 b_2 - 8 b_3 + 6 b_4$ and $b_7 =
6 b_2 - 15 b_3 + 10 b_4$. Under those conditions, one has $c_1=3a_1+b_2A$, $c_2=3a_1^2-S_{3,3,}A$, $c_3=a_1^3+S_{3,4}A$, so that Eq.\ \eqref{1}
becomes
$Z(\eta)=1+\eta\left[b_2+(b_3-3b_2)\eta+(b_4-3b_3+3b_2)\eta^2\right]/(1-\eta)^3$,
regardless of the values of $a_1$ and $A$. The aforementioned
relationships are precisely satisfied by the virial and
compressibility routes to the EOS in the PY approximation for $d=3$.
Therefore, the functional form \eqref{1} is general enough as to
include both PY EOSs, and thus also the Carnahan--Starling (CS)
EOS,\cite{CS69} given by

\beq
Z_{\text{CS}}(\eta)=\frac{1+\eta+\eta^2-\eta^3}{(1-\eta)^3},
\label{CS}
\eeq
as particular cases. Moreover, in the one-dimensional case one has
$b_j=1$, so that again the relationships are satisfied and the
resulting compressibility factor reduces to the exact EOS of the
system, namely $Z(\eta)=1/(1-\eta)$.

The numerical values of the coefficients $a_1$, $a_2$, $A$, and
$c_1$--$c_3$ obtained from the known values of the first seven
virial coefficients\cite{LKM05,CM06} (namely, $b_2=4$, $b_3=10$,
$b_4\simeq 18.364768$, $b_5\simeq 28.2245$, $b_6\simeq 39.8151$,
$b_7\simeq 53.3444$) are given in Table \ref{tab1}. The two branch
points $-(a_1\pm\sqrt{a_1^2-a_2})/a_2=-0.139234 \pm 0.702817 i$ lie
on the complex plane. Their modulus is $1/\sqrt{a_2}=0.716$ and this
is then the radius of convergence of the virial series of the EOS
\eqref{1}. While this radius is possibly an overestimate (in fact,
it is larger than the freezing density), it is not unphysical since
it is smaller than the close-packing value, in contrast to the
radius $\eta=1$ given by the PY, the CS, and the
Carnahan--Starling--Kolafa (CSK)\cite{Kolafa} EOSs, to name just a
few.

\begingroup
\squeezetable
\begin{table}
\caption{\label{tab2} Exact and predicted values of $b_8$--$b_{10}$
in the three-dimensional case}.
\begin{ruledtabular}
\begin{tabular}{ccccc}
\multirow{2}{*}{Coefficient} &
\multirow{2}{*}{Exact}&Branch-point&Rescaled expansion&Pad\'e [3/3]
\\
&&Eq.\ \protect\eqref{1}, $k=3$&Eq.\ \protect\eqref{BC}, $m=6$,
$k=3$&Eq.\ \protect\eqref{[3/3]}\\
\hline
$b_8$&$68.538$&$68.609$&$68.812$&$69.040$\\
$b_9$&$85.813$&$85.532$&$86.219$&$87.147$\\
$b_{10}$&$105.78$&$104.32$&$105.56$&$107.93$\\
\end{tabular}
\end{ruledtabular}
\end{table}
\endgroup
Table \ref{tab2} compares the known\cite{LKM05,CM06} and predicted
values of $b_8$--$b_{10}$. Apart from the values predicted by Eq.\
\eqref{1}, the table also includes the values obtained from the two
following approximate EOSs that also make use of $b_2$--$b_7$: the
rescaled virial expansion\cite{BC87}

\beq
Z=\frac{1+\sum_{n=1}^m C_n\eta^n}{(1-\eta)^k},
\label{BC}
\eeq
where $m=6$, $k=3$, and
$C_n=(-1)^n\left[\binom{k}{n}-S_{k,n+1}\right]$, with $S_{k,n}$
given by Eq.\ \eqref{8}, and the best\cite{GB08} Pad\'e approximant
[3/3] given by

\begin{figure}[h]
\includegraphics[width=1.0\columnwidth]{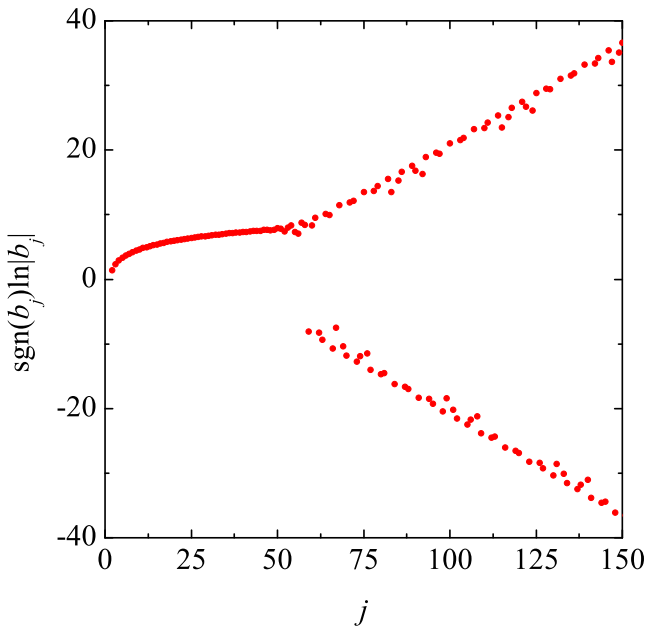}
\caption{(Color online)  Plot of  $\text{sgn}(b_j)\ln |b_j|$ for
$j\leq 150$ in the three-dimensional case.
\label{fig1}}
\end{figure}
\beq
Z=\frac{1+\sum_{n=1}^3 D_n\eta^n}{1+\sum_{n=1}^3 E_n\eta^n}.
\label{[3/3]}
\eeq
where $D_n$ and $E_n$ ($n=1,2,3$) are combinations of $b_2$--$b_7$ whose explicit
expressions may be easily obtained but will be omitted here. The
deviations from the correct ones of the  values for $b_8$--$b_{10}$
predicted by Eq.\ \eqref{1} are $0.1\%$, $0.3\%$, and $1.4\%$,
respectively. In contrast, the deviations of the values predicted by
the rescaled virial expansion and the Pad\'e approximant [3/3] are
$0.4\%$, $0.5\%$, and $0.2\%$ and $0.7\%$, $1.5\%$, and $2\%$,
respectively. Note that, in particular, the rescaled virial
expansion predicts a very accurate value for $b_{10}$, even better
than the prediction for $b_8$. At a qualitative level, an
interesting outcome of Eq.\ \eqref{1} is first that it predicts a
\emph{negative} value of a certain coefficient (specifically,
$b_{59}$) and secondly that henceforth the coefficients change sign
every 1-2 terms. Figure \ref{fig1} shows $\text{sgn}(b_j)\ln |b_j|$
for $j\leq 150$. In contrast, the rescaled virial expansion predicts
that \emph{all} the $b_j$ are positive, while the Pad\'e [3/3]
predicts positive coefficients up to $b_{56}$ and then alternating
signs for groups of $55$ consecutive coefficients.

\begin{figure}[h]
\includegraphics[width=1.0\columnwidth]{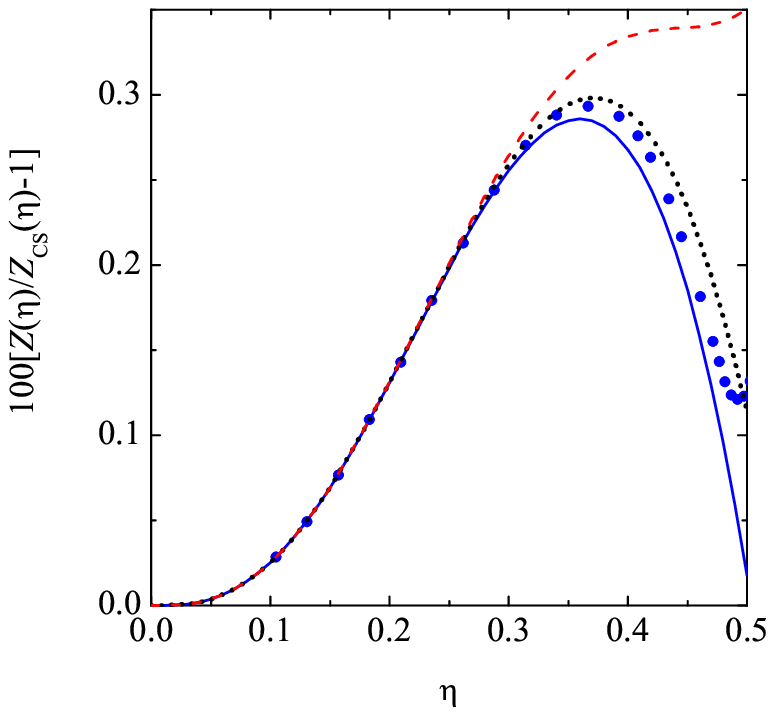}
\caption{(Color online)  Plot of $100[Z(\eta)/Z_\text{CS}(\eta)-1]$
in the three-dimensional case. Solid line: $Z(\eta)$ given by the
branch-point approximant \eqref{1} with $k=3$; dotted line:
$Z(\eta)$ given by the rescaled virial approximant \eqref{BC} with
$m=6$ and $k=3$; dashed line: $Z(\eta)$ given by the Pad\'e [3/3]
approximant \eqref{[3/3]}. The circles are simulation data from
Ref.\ \protect\onlinecite{KLM04}.
\label{fig2}}
\end{figure}
 While the comparison between the exact values of the higher reduced virial coefficients
and those that follow from the expansion of Eq.\ \eqref{1} is quite
satisfactory, one may reasonably wonder how the new EOS will perform
when compared with other accurate proposals. Figure \ref{fig2} shows
that both the branch-point approximant and the rescaled virial
expansion deviate less than $0.3\%$ from the CS values for $0\leq
\eta\leq 0.5$ and are in very good agreement with simulation
data\cite{KLM04}. The Pad\'e [3/3] does a poorer job in this
instance. Hence, the performance of the new proposal is also very
accurate over the whole fluid phase range.

\section{Other dimensionalities}
\label{sec2}
In this Section we perform a similar analysis of the use of Eq.\
\eqref{1} with $k=d$ for all the dimensionalities ($d=2,4,5,6,7,8,$
and $9$) where the first ten virial coefficients are
known.\cite{CM06,BCW08}

\begin{table*}
\caption{\label{tab3} Exact and predicted values of $b_8$--$b_{10}$
for $d=2$ and $d=4$--$9$}.
\begin{ruledtabular}
\begin{tabular}{ccccc}
\multirow{2}{*}{Coefficient} &
\multirow{2}{*}{Exact}&Branch-point&Rescaled expansion&Pad\'e [3/3]
\\
&&Eq.\ \protect\eqref{1}, $k=d$&Eq.\ \protect\eqref{BC}, $m=6$,
$k=d$&Eq.\ \protect\eqref{[3/3]}\\
\hline
\multicolumn{5}{c}{$d=2$}\\
$b_8$&$8.3191$&$8.3397$&$8.3408$&$8.3241$\\
$b_9$&$9.2721$&$9.3711$&$9.3297$&$9.3001$\\
$b_{10}$&$10.216$&$10.469$&$10.319$&$10.298$\\
\multicolumn{5}{c}{$d=4$}\\
$b_8$&$605.66$&$284.49$&$486.07$&$543.55$\\
$b_9$&$739.88$&$3339.4$&$562.33$&$605.51$\\
$b_{10}$&$1516.7$&$-5388.0$&$579.18$&$704.91$\\
\multicolumn{5}{c}{$d=5$}\\
$b_8$&$-3.0064\times 10^4$&$-3.0177\times 10^4$&$3.1662\times
10^4$&$-2.6584\times 10^4$\\
$b_9$&$3.2083\times 10^5$&$3.1961\times 10^5$&$9.4841\times
10^4$&$2.3254\times 10^5$\\
$b_{10}$&$-3.3810\times10^6$&$-2.9014\times10^6$&$2.2311\times
10^5$&$-1.6907\times10^6$\\
\multicolumn{5}{c}{$d=6$}\\
$b_8$&$-3.0752\times10^7$&$-3.0448\times10^7$&$9.4362\times
10^6$&$-2.6002\times10^7$\\
$b_9$&$7.3370\times10^8$&$7.1562\times10^8$&$3.42934\times10^7$&$5.0061\times
10^8$\\
$b_{10}$&$-1.8472\times10^{10}$&$-1.7587\times10^{10}$&$9.3201\times10^7$&$-9.5796\times10^9$\\
\multicolumn{5}{c}{$d=7$}\\
$b_8$&$-8.7684\times10^9$&$-8.6759\times10^9$&$1.3044\times10^9$&$-7.6577\times
10^9$\\
$b_9$&$4.7482\times
10^{11}$&$4.6063\times10^{11}$&$5.3315\times10^9$&$3.3839\times10^{11}$\\
$b_{10}$&$-2.7274\times10^{13}$&$-2.5651\times10^{13}$&$1.6168\times10^{10}$&$-1.4941\times10^{13}$\\
\multicolumn{5}{c}{$d=8$}\\
$b_8$&$-1.6114\times10^{12}$&$-1.5950\times10^{12}$&$1.2461\times10^{11}$&$-1.4369\times10^{12}$\\
$b_9$&$1.8713\times10^{14}$&$1.8107\times10^{14}$&$5.6751\times10^{11}$&$1.3783\times10^{14}$\\
$b_{10}$&$-2.3160\times10^{16}$&$-2.1589\times10^{16}$&$1.9031\times
10^{12}$&$-1.3220\times 10^{16}$\\
\multicolumn{5}{c}{$d=9$}\\
$b_8$&$-2.3219\times10^{14}$&$-2.2913\times10^{14}$&$9.6059\times10^{12}$&$-2.0919\times10^{14}$\\
$b_9$&$5.5879\times10^{16}$&$5.3746\times10^{16}$&$4.8352\times10^{13}$&$4.1962\times10^{16}$\\
$b_{10}$&$-1.4436\times10^{19}$&$-1.3252\times10^{19}$&$1.7789\times10^{14}$&$-8.4195\times10^{18}$\\
\end{tabular}
\end{ruledtabular}
\end{table*}

Table \ref{tab3} displays the exact and predicted values of
$b_8$--$b_{10}$ for $d=2$ and $d=4$--$9$ as given by the
branch-point approximant, the rescaled virial expansion and the
Pad\'e [3/3] EOSs.

For $d=2$ the best approximant is the Pad\'e [3/3]. The branch-point
approximant also does a good job in the case of hard disks, but the
rescaled virial approximation is slightly better (except for the
value of $b_{8}$). The case $d=4$ is somewhat peculiar because the
predictions from all the approximants are rather poor. In any event,
the Pad\'e [3/3] gives the `best' performance, followed by the
rescaled virial expansion, and finally the branch-point approximant.
This latter even `anticipates' the likely aternating character of
the series and predicts a negative value of $b_{10}$. The situation
changes for $5\leq d\leq 9$ where the performance of the rescaled
virial expansion is extremely poor and in fact it never predicts
negative coefficients, even when the exact $b_6<0$ is introduced
($d=6$, $d=7$) or the exact $b_4<0$ and $b_6<0$ are introduced
($d=8$, $d=9$). On the other hand, in these dimensionalities the
Pad\'e [3/3] predicts the right signs, while the branch-point
approximant predicts, in addition, very good values.

\begin{figure}[h]
\includegraphics[width=1.0\columnwidth]{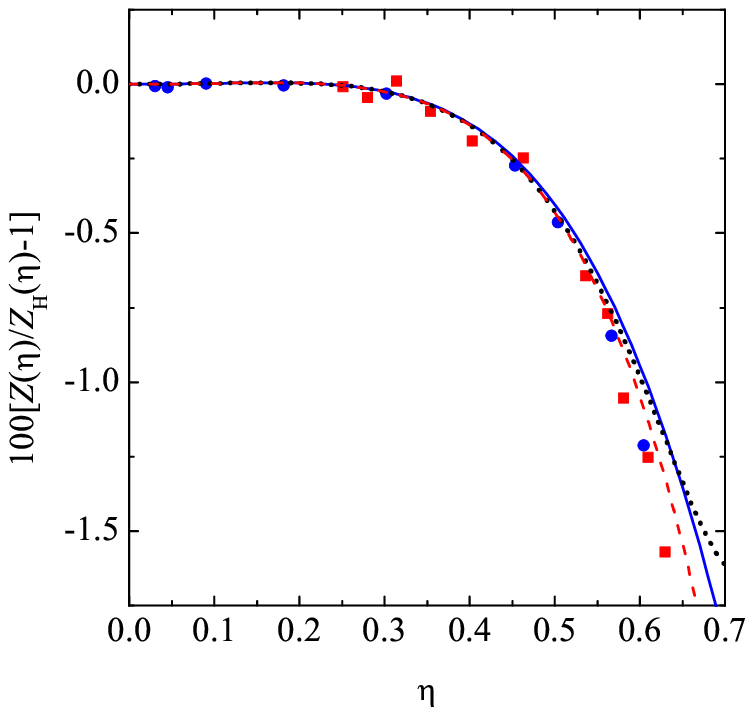}
\caption{(Color online)  Plot of $100[Z(\eta)/Z_\text{H}(\eta)-1]$
in the two-dimensional case. Solid line: $Z(\eta)$ given by the
branch-point approximant \eqref{1} with $k=2$; dotted line:
$Z(\eta)$ given by the rescaled virial approximant \eqref{BC} with
$m=6$ and $k=2$; dashed line: $Z(\eta)$ given by the Pad\'e [3/3]
approximant \eqref{[3/3]}. The circles and squares are simulation
data from Refs.\ \protect\onlinecite{EL85} and
\protect\onlinecite{L01}, respectively.
\label{fig3}}
\end{figure}
With regards to the EOS of hard disks, in Fig.\ \ref{fig3} we
compare the performance of the different approximants with respect
to the simulation data\cite{EL85,L01} in the range $0\leq \eta\leq
0.7$. Note that in this range the new proposal is able to capture
the deviations (of up to $2 \%$) of the simulation results that one
gets from the use of the reasonably accurate EOS due to Henderson,\cite{H75}
namely
\beq
Z_{\text{H}}(\eta)=\frac{1+\eta^2/8}{(1-\eta)^2}.
\label{H}
\eeq

\begin{table}
\caption{\label{tab4} Singularity closest to the origin and radius
of convergence of the virial series, as predicted by the
branch-point approximant \protect\eqref{1} with $k=d$. The radius
predicted by the PY integral equation is also included.}
\begin{ruledtabular}
\begin{tabular}{cccc}
$d$&Singularity&Radius&Radius (PY)\\
\hline
$2$&$0.3234 \pm 0.4533i$&$0.557$&$1$\\
$3$&$-0.1392 \pm 0.7028 i$&$0.716$&$1$\\
$4$&$-0.044223 \pm 0.07526i$&$0.0873$&$0.15$\\
$5$&$-0.07838$&$0.0784$&$0.057$\\
$6$&$-0.02960$&$0.0296$&$0.024$\\
$7$&$-0.01302$&$0.0130$&$0.011$\\
$8$&$-0.006062$&$0.00606$&$0.0051$\\
$9$&$-0.002925$&$0.00292$&$0.0024$\\
\end{tabular}
\end{ruledtabular}
\end{table}
Concerning the nature of the singularities in these dimensions, in
Table \ref{tab4}, we present the values of the singularity closest
to the origin and the radius of convergence of the corresponding
virial series, as predicted by the branch-point approximant
\eqref{1} with $k=d$. For comparison, the radius predicted by the PY
integral equation is also included in this table. One finds that the
new proposal predicts complex branch points for $2\leq d\leq4$.
These are precisely the cases where all the known virial
coefficients are positive. On the other hand, for $5\leq d\leq 9$
the branch point closest to the origin is a negative real value.
This agrees with the PY results, which gives some support to the
alternating series scenario. Also note that the branch-point
approximant and the PY radii of convergence tend to agree as $d$
increases.

\section{Concluding remarks}
\label{sec3}
In this paper we have introduced a new proposal for the EOS of a
three-dimensional HS fluid which is built from the knowledge of the
first seven virial coefficients and possesses two branch-point
singularities in the complex plane. Although the choice we have made
may appear to a certain extent arbitrary, it is perhaps the simplest
one embodying the PY and CS EOSs for HSs in three dimensions as well
as the exact $Z$ in $d=1$ and the PY virial EOS in $d=5$. The same
functional form was also assumed for the EOS of HS fluids in other
dimensions. For $d=3$ the new EOS predicts accurate values of the
higher virial coefficients, a radius of convergence smaller than the
close-packing value, and, irrespective of the fact that its
construction did not aim at accuracy, it is very accurate when
compared to simulation results and with other approximants involving
the same number of known virial coefficients. This last feature was
shown to be also shared by the two-dimensional case. The proposal is
also robust with respect to small ($\sim 1 \%$) deviations in the
value of the seventh virial coefficient, certainly more robust than
either the rescaled virial expansion or the Pad\'e [3/3].

Except for $d=4$ (where, as already pointed out by Clisby and
McCoy\cite{CM06} in a somewhat related context, perhaps one would
require better accuracy of the known virial coefficients), in all
other dimensionalities the branch-point approximant gives the best
overall performance with respect to the prediction of the known
virial coefficients. In particular, the rescaled virial expansion is
unable to predict even the signs of known virial coefficients for
$d\geq 5$ and the Pad\'e [3/3], although correctly capturing these
signs, leads to higher deviations. This of course constitutes no
proof that the true EOS of HS systems should include a branch-point
singularity, but the evidence provided here is at least consistent
with it.

On a related vein, the new EOS for HSs in $d=3$ also leads to an
alternating virial series, with $b_{59}$ being the first negative
reduced virial coefficient. Given the difficulty of computing exact
high order virial coefficients, it is unlikely that the alternating
series scenario for HSs in three dimensions may be confirmed in the
near future. However, in view of the present results and those
obtained in higher dimensions,\cite{CM06,RMHS08,AKV08} it certainly
gets reinforced.

One can reasonably wonder whether the present approach to construct
the EOS of HS systems using a number of known virial coefficients
may be cast in a systematic way. While the answer is certainly not
unique, the following constitutes a \emph{possible} generalization.
We rewrite the compressibility factor as

\beq
Z(\eta)=1+\frac{1+\sum_{n=1}^{2N+1}c_n\eta^n
-\left(1+2a_1\eta+a_2\eta^2\right)^{N+1/2}}{A(1-\eta)^k},
\label{2}
\eeq
where taking $N=1$ corresponds to Eq.\ \eqref{1}. In the
three-dimensional case ($k=d=3$), the approximant with $N=0$ (which
amounts to including only the first five virial coefficients)
predicts $b_6$--$b_{10}$ with deviations equal to $0.54\%$,
$0.69\%$, $1.2\%$, $3\%$, and $5\%$, respectively. On the other
hand, the approximant with $N=2$ predicts $b_{10}$ with a $1.2\%$
deviation. Therefore, although there is certainly an improvement in
the prediction of $b_{10}$ on going from $N=0$ to $N=2$, a
reasonable compromise between simplicity, generality, and accuracy
seems to suggest that the choice $N=1$ is the most adequate.

Finally, it should be pointed out that if instead of choosing $k=d$
as we have done in this paper, a different $k$ is picked (say $k=3$
for all $d$) we find slight variations in the numerical predictions
but the overall picture remains unaltered.

\begin{acknowledgments}
We are grateful to an anonymous referee for useful suggestions. This work has been supported by the Ministerio de Educaci\'on y
Ciencia (Spain) through Grant No.\ FIS2007–60977 (partially financed
by FEDER funds) and by the Junta de Extremadura through Grant No.\
GRU09038.
\end{acknowledgments}

\end{document}